\documentclass[9pt,twocolumn,twoside]{osajnl}

\usepackage{epstopdf}

\journal{josaa} 
\setboolean{shortarticle}{false} 

\title{Amplitude and phase beam shaping for the highest sensitivity in side-wall angle detection}

\author[1,*]{Luca Cisotto}
\author{Paul H. Urbach}

\affil{Optics Research Group, Faculty of Applied Sciences, Delft University of Technology, Van der Waalsweg 8, 2628 CH Delft, The Netherlands}
\affil[*]{Corresponding author: l.cisotto@tudelft.nl}

\dates{Compiled \today}

\ociscodes{}

\doi{\url{http://dx.doi.org/10.1364/ao.XX.XXXXXX}}

\def\Xint#1{\mathchoice
{\XXint\displaystyle\textstyle{#1}}%
{\XXint\textstyle\scriptstyle{#1}}%
{\XXint\scriptstyle\scriptscriptstyle{#1}}%
{\XXint\scriptscriptstyle\scriptscriptstyle{#1}}%
\!\int}
\def\XXint#1#2#3{{\setbox0=\hbox{$#1{#2#3}{\int}$}
\vcenter{\hbox{$#2#3$}}\kern-.5\wd0}}

\def\dashint{\Xint-}

\begin{abstract}
In optical metrology, grating-like structures are used as tools to evaluate the performance of lithographic techniques. In particular, several shape parameters characterize those structures. One of them, termed side-wall angle, suffers from a considerable high error estimation. Using mathematical optimization, we investigated whether a properly shaped beam could increase the ability to detect tiny changes of this angle in the case of a cliff-like structure. This paper describes the theoretical formulation used to calculate the optimized beam and compares its performance with the case of a plane wave. The results indicate that the sensitivity increases by several folds by using the optimized solution. Still, such an optimization process needs to be extended to the more general vectorial case.
\end{abstract}

\setboolean{displaycopyright}{true}

\begin{document}

\maketitle
\thispagestyle{fancy}
\ifthenelse{\boolean{shortarticle}}{\abscontent}{}

\section{Introduction}

Nowadays, the performance increase of electronic devices is mainly achieved by two highly related means: a bigger density of electrical components (such as transistors) and a miniaturization of those. It is therefore important, in order to obtain the desired characteristics and performances, to implement a reliable and (possibly) fast process control. This is achieved by printing special targets on the wafer, such as gratings, which are measured in order to adjust dose, exposure time, overlay/alignment and other relevant process parameters of the photo-lithographic machine \cite{Edwards,Vogel}. Currently, one of the important tasks of this methodology is to quantify the value of the so called side-wall angle (SWA) with high precision (for example tenth or hundredth of a degree). In the past years, several different techniques have been proposed to obtain a direct measurement of this quantity, examples can be Atomic Force Microscopy (AFM) and Scanning Electron Microscopy (SEM); nevertheless, technological innovation soon demanded more accurate and reliable metrology tools. For this reason, researchers developed numerous variations of the aforementioned techniques, such as a tilt-scanning method for AFM \cite{Murayama}, Critical Dimension AFM \cite{Dai} and Critical Dimension SEM \cite{Bingham}. These measuring tool all have the great feature of being able to perform non-destructive measurement, nevertheless they all have different drawbacks (for instance the Rayleigh diffraction limit for the AFM) affecting their performances. To overcome these problems, El Gawhary et al. \cite{Omar, Nitish, Roy} developed a completely new approach, Coherent Fourier Scatterometry, which they use to reconstruct the profile of a grating; nonetheless, even in this case, the uncertainty associated to the side-wall angle is still much higher than the error affecting other shape parameters (e.g. height and middle critical dimension).

Phase and amplitude control of light open many new applications in the field of optics. In adaptive optics, for example, modulated light can remove aberrations introduced by very different optical systems such as the human eye or the atmosphere \cite{Tyson, Rai}. In biology, a properly shaped beam can be used to efficiently trap different types of particles \cite{Leach, Grier, Eriksson}. Another area where beam shaping plays a key role is digital holography; in this field, an example of the importance of light modulation can be holographic data storage \cite{Heanue}. Another innovative application of beam shaping is the focusing of light onto strongly scattering materials \cite{Vellekoop1, Vellekoop2}. The scientific progress in the field of light modulation has been so fast that we can already consider changing the amplitude, phase as well as the polarization of a beam. The great diffusion of modulating devices, such as Spatial Light Modulators (SLM), have made possible to individually tune each of the just named properties of light. Particularly, several authors claim to be able to create all the possible state of polarizations within one single SLM, alongside with changes in the amplitude and phase \cite{Chen, Han}.

In this paper, we optimize the illumination to increase the sensitivity (here intended as ability to sense changes of a specific parameter) in the detection of the side-wall angle (SWA) of a cliff-like structure. We assume that a cylindrical lens is used, which gives a focused spot that is independent of the $y$-component; the structure we intend to study constitutes a phase object, meaning that absorption is not considered in our model and only the phase of the beam is affected by the interaction with the sample. The paper is organized as follows: in Section~\ref{sec:opt_prob} we present the optimization problem used to calculated the optimum field for a given side-wall angle; in Section~\ref{sec:lagrange_prob} this problem is expressed in terms of a Lagrange multiplier rule, which turns into an eigenvalue problem; in Section~\ref{sec:kernel_K} we derive a closed formula for the kernel $\mathcal{K}$, from which we will calculate the eigenvalues of the system; Section~\ref{sec:discretiz} deals with the description of the computational techniques we used to calculate the optimum field; in Section~\ref{sec:results} we present the main results obtained from the optimization algorithm and we compare them with the case of a plane wave illumination; Section~\ref{sec:conclusion} summarizes the main findings.

\section{THE OPTIMIZATION PROBLEM}\label{sec:opt_prob}

\begin{figure}[htbp]
\centering
\fbox{\includegraphics[scale=0.65]{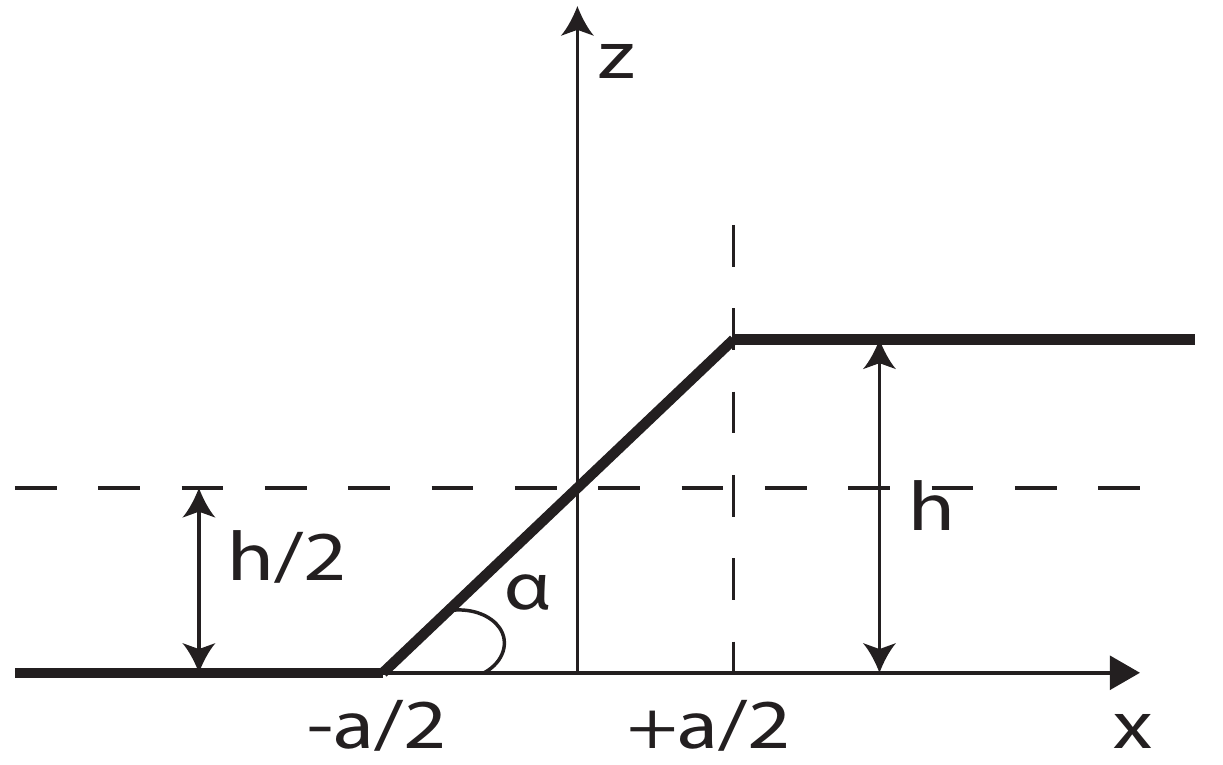}}
\caption{The height $h$ and side-wall angle $\alpha$ are the parameters that we will vary. The quantity $a$ is automatically derived from the other two variables with Eq.~(\ref{eq:equation_for_a}).}
\label{fig:slope_function}
\end{figure}

In this section, we present the theoretical derivation of the optimized pupil field that is, for a given power, most sensitive to a change in the side-wall angle in the scalar regime. The structure we consider has a cliff like shape and it is characterized by a side-wall angle (SWA) $\alpha$ and height $h$. We center it in a $(x,y)$ reference system, such that the slope extends in the range $-a/2<x<+a/2$, with $a$ being the distance from the $y$-axis (see Fig.~\ref{fig:slope_function}). The surface can therefore mathematically be expressed by the function $g_\alpha(x)$:
\begin{equation}
g_\alpha(x):=
\begin{cases}
0, & \text{if}\quad x\leqslant -\frac{a}{2}\\
x\tan(\alpha)+\frac{h}{2}, &  \text{if}\quad -\frac{a}{2}<x<\frac{a}{2}\\
h, & \text{if}\quad x\geqslant\frac{a}{2}
\end{cases}
\end{equation}

Throughout the mathematical derivation, we will use $\alpha$ and $h$ as variables to describe the cliff, since the horizontal length $a$ can be directly obtained from the formula:
\begin{equation}
a = \frac{h}{\tan(\alpha)} \qquad 0<\alpha<\frac{\pi}{2}.\label{eq:equation_for_a}
\end{equation}
Moreover, we will treat the sample described by the function $g_\alpha(x)$ as a pure phase object, meaning that we don't consider absorption in the calculations that will follow. In view of this hypothesis, the reflection function of the structure becomes:
\begin{equation}\label{eq:r_alpha}
r_\alpha(x)=
\begin{cases}
\exp{(2i k h)}, & \text{if}\quad x\leqslant -\frac{a}{2}\\
\exp{\left[2i k \left(\frac{h}{2}-x\tan(\alpha)\right)\right]}, &  \text{if}\quad -\frac{a}{2}<x<\frac{a}{2}\\
1, & \text{if}\quad x\geqslant\frac{a}{2}
\end{cases}
\end{equation}
where $k$ is the wave number $k=2\pi/\lambda$.

Let $U^i(x)$ be a cylindrical focused field incident on the sample described by the reflection function $r_\alpha(x)$. We assume that the reflected field $U^r_\alpha(x)$ is given by:
\begin{equation}
U^r_\alpha(x) = U^i(x)\,r_\alpha(x).
\end{equation}
Suppose that the focused field $U^i(x)$ is generated by a cylindrical microscope objective of a given numerical aperture (NA). If we think of an experimental configuration in which the reflected light is collected by the same objective and then analyzed with a CCD camera, the intensity captured by this detector is simply the absolute square of the Fourier transform $\mathcal{F}$ of the reflected field $U^r_\alpha(x)$:
\begin{equation}
I^{\text{out}}_\alpha (\xi) = |\mathcal{F}(U^r_\alpha)(\xi)|^2 = |\mathcal{F}(U^i\,r_\alpha)(\xi)|^2 \qquad |\xi|\leq \frac{\text{NA}}{\lambda},
\end{equation}
where the newly introduced coordinate $\xi$ refers to the CCD plane.
Recalling the properties of the Fourier transform, we can write:
\begin{equation}
\mathcal{F}(U^i\,r_\alpha)(\xi) = \mathcal{F}(U^i)(\xi)*\mathcal{F}(r_\alpha)(\xi) = A^i(\xi)*\mathcal{F}(r_\alpha)(\xi),\label{eq:simplify_Ai}
\end{equation}
where the convolution operation is expressed by the symbol $*$ and, in the last step of Eq.~(\ref{eq:simplify_Ai}), we have introduced the variable $A^i(\xi)$ for a more concise notation. Throughout the mathematical derivation, $A^i(\xi)$ represents the complex amplitude of the expansion into plane waves of the incident field. These complex amplitudes are the variables of the optimization problem.

In order to find the input field $A^i(\xi)$ which provides the highest sensitivity to the side-wall angle, we maximize the change, averaged over the microscope objective pupil, of the output intensity $I^{\text{out}}_\alpha(\xi)$. Hence we seek $A^i(\xi)$ such that:
\begin{align}
G(A^i) &= \int_{-\frac{\text{NA}}{\lambda}}^{+\frac{\text{NA}}{\lambda}} \frac{d}{d\alpha}I^{\text{out}}_\alpha (\xi)\,d\xi\notag \\
&= \int_{-\frac{\text{NA}}{\lambda}}^{+\frac{\text{NA}}{\lambda}} \frac{d}{d\alpha} |A^i*\mathcal{F}(r_\alpha)|^2(\xi)\,d\xi\notag \\
&= 2\Re\int_{-\frac{\text{NA}}{\lambda}}^{+\frac{\text{NA}}{\lambda}} \left(A^i*\frac{d\mathcal{F}(r_\alpha)}{d\alpha}\right)\left({A^i}*\mathcal{F}(r_\alpha)\right)^{*}\,d\xi,\label{eq:functional_G}
\end{align}
is either strongest positive or strongest negative for the given power:
\begin{equation}
P(A^i)=\int_{-\frac{\text{NA}}{\lambda}}^{+\frac{\text{NA}}{\lambda}} A^i(\xi)\,{A^i}^*(\xi)\,d\xi = 1\label{eq:power_1}.
\end{equation}
The superscript $*$ indicates the complex conjugate and we used Eq.~(\ref{eq:simplify_Ai}) between the first and second step in the right-hand side of Eq.~(\ref{eq:functional_G}). 

\section{THE LAGRANGE MULTIPLIER RULE}\label{sec:lagrange_prob}

A useful tool to find a solution to this problem is provided by the Lagrange multiplier rule for inequality constraints \cite{Luenberger} (also known as Kuhn-Tucker’s theorem); according to this rule there exists a Lagrange multiplier $\Lambda$ such that, if $A^i$ is the optimum field, the following equation needs to be satisfied for all complex $B^i$:
\begin{equation}
\delta G(A^i)(B^i) - \Lambda\, \delta P(A^i)(B^i) = 0\label{eq:lagrange_eq}
\end{equation}
where $\delta G(A^i)(B^i)$ and $\delta P(A^i)(B^i)$ represent the Fr\'{e}chet derivative of the functionals $G(A^i)$ and $P(A^i)$ in the direction of $B^i$. If $\Lambda>0$, then the solution is such that $G(A^i)$ is strongest positive, whereas if $\Lambda<0$, $G(A^i)$ is strongest negative. We have:
\begin{align}\label{eq:freshcet_G}
\delta G(A^i)(B^i) &= \begin{aligned}[t]
					&2\Re\int_{-\frac{\text{NA}}{\lambda}}^{+\frac{\text{NA}}{\lambda}} \biggl[{A^i}^{*}*\mathcal{F}(r_\alpha)\;B^i*\frac{d\mathcal{F}(r_\alpha)}{d\alpha}+\notag\\
					&+{A^i}^{*}*\frac{d\mathcal{F}(r_\alpha)}{d\alpha}\;B^i*\mathcal{F}(r_\alpha)\biggr]\,d\xi\notag
					  \end{aligned}\\
&= \begin{aligned}[t]
		&2\Re\biggl\{\int_{-\frac{\text{NA}}{\lambda}}^{+\frac{\text{NA}}{\lambda}} \left[{A^i}^{*}*\mathcal{F}(r_\alpha)\right](\xi)\times\notag\\
		&\times\int_{-\frac{\text{NA}}{\lambda}}^{+\frac{\text{NA}}{\lambda}}\frac{d\mathcal{F}(r_\alpha)}{d\alpha}(\xi-\xi')B^i(\xi')d\xi'\,d\xi +\\
		&+\int_{-\frac{\text{NA}}{\lambda}}^{+\frac{\text{NA}}{\lambda}}\left[{A^i}^{*}*\frac{d\mathcal{F}(r_\alpha)}{d\alpha}\right](\xi)\times\\
		&\times\int_{-\frac{\text{NA}}{\lambda}}^{+\frac{\text{NA}}{\lambda}}\mathcal{F}(r_\alpha)(\xi-\xi') B^i(\xi')\,d\xi'\,d\xi\biggr\}
	\end{aligned}\\
&= \begin{aligned}[t]
		&2\Re\int_{-\frac{\text{NA}}{\lambda}}^{+\frac{\text{NA}}{\lambda}}\int_{-\frac{\text{NA}}{\lambda}}^{+\frac{\text{NA}}{\lambda}} \mathcal{K}(\xi,\xi')\,A^i(\xi')\,d\xi'B^i(\xi)\,d\xi.
	\end{aligned}
\end{align}
Where the Kernel $\mathcal{K}(\xi,\xi')$ is symmetric and can be expressed as:
\begin{equation}
\mathcal{K}(\xi,\xi')=\mathcal{H}(\xi,\xi')+\mathcal{H}(\xi',\xi)\label{eq:kernel_symmetric}
\end{equation}
with:
\begin{equation}
\mathcal{H}(\xi,\xi') = \int_{-\frac{\text{NA}}{\lambda}}^{+\frac{\text{NA}}{\lambda}} \frac{d\mathcal{F}(r_\alpha)}{d\alpha}(\xi''-\xi)\,\mathcal{F}(r_\alpha)(\xi''-\xi') d\xi''\label{eq:H_term}
\end{equation}
Since $\mathcal{K}(\xi,\xi')=\mathcal{K}(\xi',\xi)$ and $\mathcal{K}$ is real valued (see Eqs.~(\ref{eq:f_r_alpha}) and~(\ref{eq:der_f_r_alpha}) in Section~\ref{sec:kernel_K}), the integral operator is self-adjoint and therefore there exists an orthogonal basis of $L^2(\mathbb{R})$ of eigenfunctions and all eigenvalues are real.
\newline The Fr\'{e}chet derivative of the power $P(A^i)$ in the direction of $B^i$ is given by: 
\begin{equation}
\delta P(A^i)(B^i) = 2\Re\int_{-\frac{\text{NA}}{\lambda}}^{+\frac{\text{NA}}{\lambda}} {A^i}^{*}*B^i\,d\xi.\label{eq:freshcet_P}
\end{equation}
Hence Eqs.~(\ref{eq:lagrange_eq}),~(\ref{eq:freshcet_G}) and ~(\ref{eq:freshcet_P}) imply:
\begin{equation}
\int_{-\frac{\text{NA}}{\lambda}}^{+\frac{\text{NA}}{\lambda}}\int_{-\frac{\text{NA}}{\lambda}}^{+\frac{\text{NA}}{\lambda}} \left[\mathcal{K}(\xi,\xi')\,A^i(\xi') - \Lambda A^i(\xi)\right]\;d\xi\,d\xi'\label{eq:eigen_eq2}
\end{equation}

\section{COMPUTATION OF THE KERNEL \texorpdfstring{$\mathcal{K}$}{K}}\label{sec:kernel_K}

The Fourier transform of the reflection function $r_\alpha$ of Eq.~(\ref{eq:r_alpha}), is given by the distribution:
\begin{align}
\mathcal{F}(r_\alpha)(\xi) = &-\text{PV}\frac{\sin\left(k h + \pi h\frac{\xi}{\tan\alpha}\right)}{\pi\xi}+\delta(\xi)\cos(k h)+\notag\\
&+\frac{\sin\left(k h+\pi h \frac{\xi}{\tan\alpha}\right)}{k\tan\alpha+\pi\xi}\label{eq:f_r_alpha}
\end{align}
where $\delta(\xi)$ is the Dirac's delta function and $\text{PV}$ indicates the generalized function (distribution) given by the Cauchy Principal Value (see Eq.~(\ref{eq:PV}) for the more complete treatment), i.e. for any smooth test function $\phi(\xi)$ we have:
\begin{align}
\text{PV} \int \frac{\sin\left(k h + \pi h\frac{\xi}{\tan\alpha}\right)}{\pi\xi} \phi(\xi) d\xi &= \frac{1}{\pi} \lim_{\epsilon\rightarrow 0^+} \int_{\Re\setminus[-\epsilon,+\epsilon]} \frac{\psi(\xi)}{\xi} d\xi\\
&= \frac{1}{\pi} \int_0^{+\infty} \frac{\psi(\xi)-\psi(-\xi)}{\xi} d\xi
\end{align}
where, in the second step, we included the sine function in the expression of $\psi(\xi)$. Similarly, the derivative with respect to $\alpha$ of the just calculated Fourier transform, is given by:
\begin{align}
\frac{d \mathcal{F}(r_\alpha)}{d\alpha}(\xi) = &\frac{k h \cos\left(kh+\pi h \frac{\xi}{\tan\alpha}\right)}{k\sin^2\alpha+\frac{\pi\xi}{2}\sin(2\alpha)}+\notag\\
&-\frac{k \sin\left(kh+\pi h \frac{\xi}{\tan\alpha}\right)}{k^2\sin^2\alpha+\pi^2\xi^2\cos^2\alpha+k\pi\xi\sin(2\alpha)}\label{eq:der_f_r_alpha}
\end{align}
note that $\mathcal{F}(r_\alpha)$ and $d\mathcal{F}(r_\alpha)/d\alpha$ are both real valued functions.
It is possible to calculate a closed formula for the function $\mathcal{H}(\xi,\xi')$ given in Eq.~(\ref{eq:H_term}) by using the explicit expressions for $\mathcal{F}(r_\alpha)$ and $d\mathcal{F}(r_\alpha)/d\alpha$ in Eqs.~(\ref{eq:f_r_alpha}) and~(\ref{eq:der_f_r_alpha}), respectively. Furthermore, we need to pay particular attention to the integration containing the Cauchy Principal Value. In fact, we write the term $\mathcal{H}(\xi,\xi')$ in Eq.~(\ref{eq:H_term}) as:
\begin{align}
\mathcal{H}(\xi,\xi') = &-\frac{1}{\pi}\dashint_{-\frac{\text{NA}}{\lambda}}^{+\frac{\text{NA}}{\lambda}} \biggl\{\frac{1}{\xi''-\xi'}\sin\left[k h + \pi h\frac{(\xi''-\xi')}{\tan\alpha}\right]\times\notag\\
&\times\frac{d\mathcal{F}(r_\alpha)}{d\alpha}(\xi''-\xi)\biggr\}d\xi'' + \cos(kh) \frac{d\mathcal{F}(r_\alpha)}{d\alpha}(\xi'-\xi) +\notag\\
&+ \int_{-\frac{\text{NA}}{\lambda}}^{+\frac{\text{NA}}{\lambda}} \frac{d\mathcal{F}(r_\alpha)}{d\alpha}(\xi''-\xi)\,\frac{\sin\left[k h+\pi h \frac{(\xi''-\xi')}{\tan\alpha}\right]}{k\tan\alpha+\pi(\xi''-\xi')}\,d\xi''.\label{eq:int_H_PV}
\end{align}
The first integral in the right-hand side of Eq.~(\ref{eq:int_H_PV}) requires careful examination. To calculate it, we define a new function $w(\xi'')$ containing part of the integrand:
\begin{equation}
w(\xi'') = \sin\left[k h + \pi h\frac{(\xi''-\xi')}{\tan\alpha}\right]\,\frac{d\mathcal{F}(r_\alpha)}{d\alpha}(\xi''-\xi)\label{eq:w_sec}
\end{equation}
The Cauchy Principal Value integral can now be written as:
\begin{align}
I_{PV}(\xi') &= -\frac{1}{\pi}\dashint_{-\frac{\text{NA}}{\lambda}}^{+\frac{\text{NA}}{\lambda}} \frac{w(\xi'')}{\xi''-\xi'} d\xi''\notag\\
&= -\frac{1}{\pi}\lim_{\epsilon\rightarrow 0}\biggl[\left(\int_{-\frac{\text{NA}}{\lambda}}^{\xi'-\epsilon}+\int_{\xi'+\epsilon}^{+\frac{\text{NA}}{\lambda}}\right) \frac{w(\xi'')-w(\xi')}{\xi''-\xi'}\,d\xi'' +\notag\\
&+ w(\xi')\left(\int_{-\frac{\text{NA}}{\lambda}}^{\xi'-\epsilon}+\int_{\xi'+\epsilon}^{+\frac{\text{NA}}{\lambda}}\right) \frac{1}{\xi''-\xi'} \,d\xi''\biggr].\label{eq:expansion_PV_int_1}
\end{align}
Since $w(\xi'')$ is differentiable, the first integral in the right-hand side of Eq.~(\ref{eq:expansion_PV_int_1}) is not singular; the second integral is singular, but it's simpler to calculate than the integral we started with, and can be estimated with a limit process:
\begin{align}
I_{PV}(\xi') &= -\frac{1}{\pi}\biggl[\int_{-\frac{\text{NA}}{\lambda}}^{+\frac{\text{NA}}{\lambda}} \frac{w(\xi'')-w(\xi')}{\xi''-\xi'}\,d\xi'' +\notag\\
&+ w(\xi')\lim_{\epsilon\rightarrow 0} \left(\int_{-\frac{\text{NA}}{\lambda}}^{\xi'-\epsilon}+\int_{\xi'+\epsilon}^{+\frac{\text{NA}}{\lambda}}\right) \frac{1}{\xi''-\xi'}\,d\xi''\biggr]\notag\\
\label{eq:expansion_PV_int_2}
&= -\frac{1}{\pi}\biggl[\int_{-\frac{\text{NA}}{\lambda}}^{+\frac{\text{NA}}{\lambda}} \frac{w(\xi'')-w(\xi')}{\xi''-\xi'}\,d\xi'' + \notag\\
&+ w(\xi')\ln\left|\frac{\frac{\text{NA}}{\lambda}-\xi'}{\frac{\text{NA}}{\lambda}+\xi'}\right|\biggr].
\end{align}
Finally, the expression of $\mathcal{H}(\xi,\xi')$:
\begin{align}
\mathcal{H}(\xi,\xi') = &-\frac{1}{\pi}\int_{-\frac{\text{NA}}{\lambda}}^{+\frac{\text{NA}}{\lambda}} \frac{w(\xi'')-w(\xi')}{\xi''-\xi'}	\,d\xi'' +\notag\\
&- \frac{1}{\pi}\ln\left|\frac{\frac{\text{NA}}{\lambda}-\xi'}{\frac{\text{NA}}{\lambda}+\xi'}\right|\,\sin\left(k h \right)\,\frac{d\mathcal{F}(r_\alpha)}{d\alpha}(\xi'-\xi) +\notag\\
&+\cos(kh) \frac{d\mathcal{F}(r_\alpha)}{d\alpha}(\xi'-\xi) +\notag\\
&+ \int_{-\frac{\text{NA}}{\lambda}}^{+\frac{\text{NA}}{\lambda}} \frac{d\mathcal{F}(r_\alpha)}{d\alpha}(\xi''-\xi)\,\frac{\sin\left[k h+\pi h \frac{(\xi''-\xi')}{\tan\alpha}\right]}{k\tan\alpha+\pi(\xi''-\xi')}\,d\xi'',\label{eq:expression_H_limit}
\end{align}
where $w(\xi'')$ is given by~(\ref{eq:w_sec}). Note that for $\xi''=\xi'$:
\begin{equation}
w(\xi'')\bigg|_{\xi''=\xi'} = \sin\left(k h \right)\,\frac{d\mathcal{F}(r_\alpha)}{d\alpha}(\xi'-\xi),
\end{equation}
At this point, if we substitute Eqs.~(\ref{eq:freshcet_G}) and~(\ref{eq:freshcet_P}) into Eq.~(\ref{eq:lagrange_eq}) and carry on the algebraic calculations, we will be left with the final expression of the Lagrange multiplier problem we need to solve in order to find the optimized input field:
\begin{equation}
\int_{-\frac{\text{NA}}{\lambda}}^{+\frac{\text{NA}}{\lambda}} \mathcal{K}(\xi,\xi')\,A^i(\xi')\,d\xi' = \Lambda A^i(\xi)\,d\xi'\label{eq:eigen_eq}
\end{equation}
this integral equation if often termed in mathematical literature as second order Fredholm equation.
The solution of the optimization problem is the properly normalized eigenfunction corresponding to the strongest positive, or strongest negative,  eigenvalue $\Lambda$.

\section{DISCRETIZATION}\label{sec:discretiz}

To be able to solve Eq.~(\ref{eq:eigen_eq}) we need to use numerical computation, since this equation cannot, in general, be calculated analytically. We used Gauss integration with the formalism described in \cite{Gil}. Gauss integration is well suited for integrals with integration domain in the range $[-1,+1]$, we therefore normalized the end points of the integration interval in the left-hand side of Eq.~(\ref{eq:eigen_eq}); setting $\xi'=\frac{\text{NA}}{\lambda}\,\xi''$ we get:
\begin{equation}
\int_{-1}^{1} \mathcal{K}\left(\xi,\frac{\text{NA}}{\lambda}\xi''\right)\,A^i\left(\frac{\text{NA}}{\lambda}\xi''\right) d\xi'' = \Lambda' A^i(\xi)\label{eq:gauss_integral_equation}
\end{equation}
where $\Lambda'=\Lambda\,\lambda/\text{NA}$. Since the integration interval of Eq.~(\ref{eq:gauss_integral_equation}) runs now from $-1$ to $1$, we can apply the Gaussian quadrature rule. The first step is to discretize the variable $\xi''$ into $N$ points $-1={\xi}_1''<{\xi}_2''<\ldots<{\xi}_N''=+1$ with Gaussian weight $w_i$ corresponding to ${\xi}_i''$, and obtain:
\begin{equation}
\sum_{n=1}^N w_n\,\mathcal{K}(\xi,\beta {\xi}_n'')\,A^i(\beta {\xi}_n'')=\Lambda'\,A^i(\xi)
\end{equation}
To proceed, we discretize the variable $\xi$ into $N$ points as well, therefore:
\begin{equation}
\sum_{n=1}^N w_n\,\mathcal{K}({\xi}_m,\beta {\xi}_n'')\,A^i(\beta {\xi}_n'')=\Lambda'\,A^i({\xi}_m)\qquad m=1,2,\ldots,N\label{eq:system}
\end{equation}
The system~(\ref{eq:system}) represents a set of $N$ equations (coming from the discretization of the variable $\xi$) and each of these equations contains a sum over $N$ terms; it is therefore more intuitive to write this system in matrix representation by setting $\mathbf{K} = \mathcal{K}({\xi}_m,\beta {\xi}_n'')$, $\mathbf{w} = \text{diag} (w_1,\ldots,w_N)$ and $\mathbf{A^i} = [A^i({\xi}_1),A^i({\xi}_2),\ldots,A^i({\xi}_N)]^T$. Hence we need to solve the matrix eigenvalue problem given by the following equation:
\begin{equation}
\mathbf{K}\mathbf{w}\,\mathbf{A^i}=\Lambda'\,\mathbf{A^i}
\end{equation}
which, when solved, yields $N$ eigenvalues $\Lambda_m'$ and the associated eigenfunctions $A_m^i$ ($m = 1,2,\ldots,N$).

\section{Results}\label{sec:results}

In this section, we will plot the behavior of the optimized field $A^i(\xi)$ for some values of the slope angle $\alpha$, the height $h$ and the numerical aperture $\text{NA}$ of the system. Furthermore, we will confront the sensitivity of the optimized field with that of a plane wave. In this way we can compare the performance of a properly shaped beam with the most used and general type of illumination. For the optimum input field, the sensitivity is simply given by the strongest positive (or strongest negative) eigenvalue $\Lambda$, whereas for a plane wave illumination we need to calculate the sensitivity $G(A^i)$ from Eq.~(\ref{eq:functional_G}), by substituting for $A^i(\xi)$ the expression of a plane wave. We have:
\begin{align}\label{eq:sens_plane_int}
G(A^i_{\text{PW}}) &= \frac{1}{2}\, \delta G(A^i_{\text{PW}})(A^i_{\text{PW}})\notag\\
&=|A^i_0|^2\Re\int_{-\frac{\text{NA}}{\lambda}}^{+\frac{\text{NA}}{\lambda}}\int_{-\frac{\text{NA}}{\lambda}}^{+\frac{\text{NA}}{\lambda}} \mathcal{K}(\xi,\xi')\,d\xi\,d\xi'
\end{align}
where $A^i_0$ represents the amplitude of the plane wave. Hence, it is possible to calculate the sensitivity of the plane wave case by simply integrating the kernel $\mathcal{K}(\xi,\xi')$. The amplitude $A^i_0$ of the plane wave has to be normalized to 1 using Eq.~(\ref{eq:power_1}), which gives $|A^i_0|^2=\lambda/(2\,\text{NA}) $. All the plots we will present have been obtained by considering the input illumination in the visible part of the spectrum ($\lambda=633nm$).

\begin{figure}[htbp]
\centering
\fbox{\includegraphics[scale=0.35]{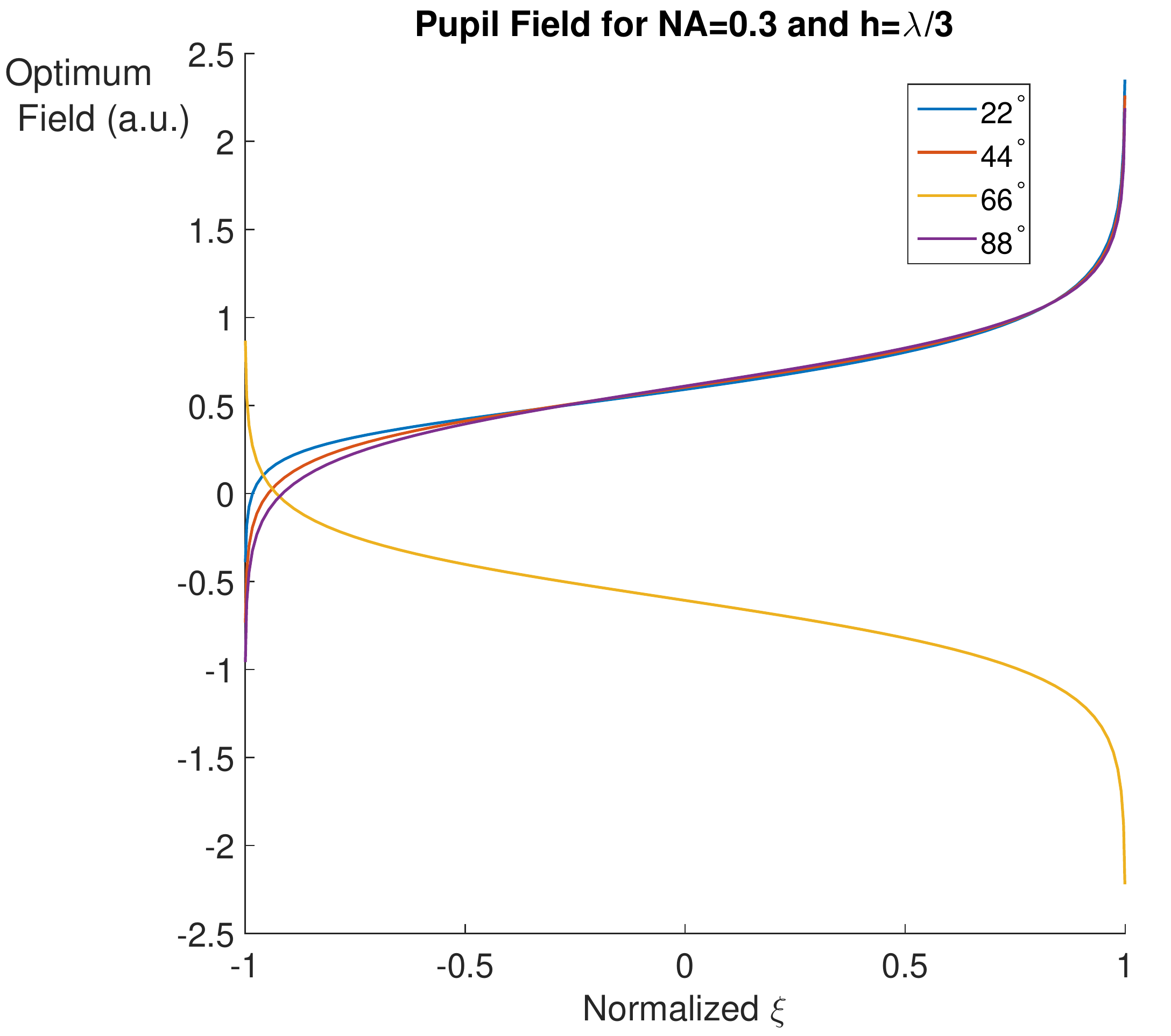}}
\caption{The optimum field changes when the slope angle changes, in this case the numerical aperture and the height of the cliff are do not vary.}
\label{fig:field_varyangle}
\end{figure}

\begin{figure}[htbp]
\centering
\fbox{\includegraphics[scale=0.35]{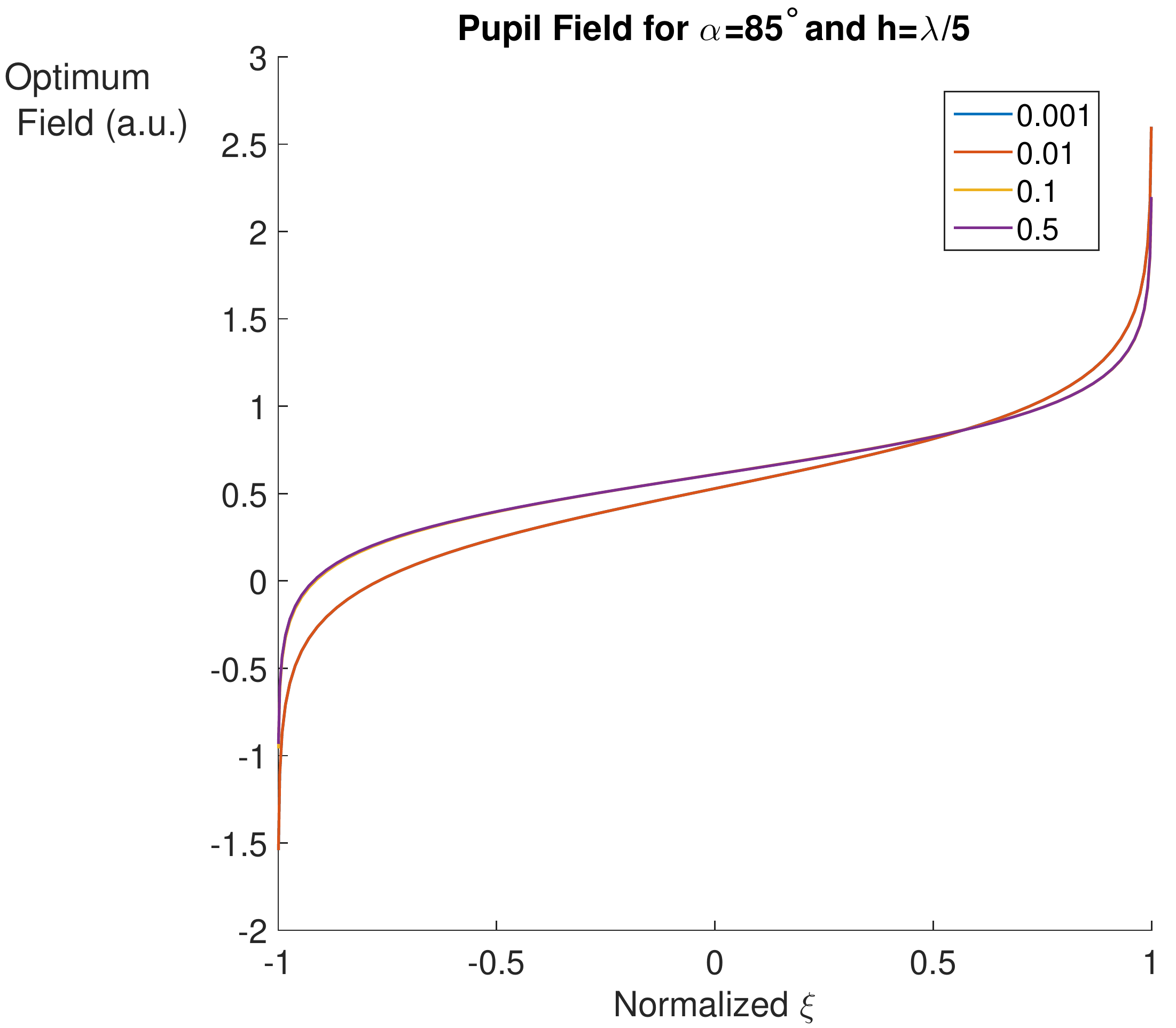}}
\caption{Changing the value of the numerical aperture of the system influences the behavior of the optimum field.}
\label{fig:field_varyNA}
\end{figure}

Figure~\ref{fig:field_varyangle} shows the profiles of the optimized input field as a function of the coordinate $\xi$ (normalized to $1$), for $\alpha=22^\circ,44^\circ,66^\circ,88^\circ$, $h=\lambda/3$, $\text{NA}=0.3$. As we already pointed out in Section~\ref{sec:lagrange_prob}, the kernel $\mathcal{K}(\xi,\xi')$ of the eigenvalue equation we need to solve, is real valued and symmetric, therefore the eigenfunctions are real valued and the phase of the optimized beam is thus $0$ or $\pi$. The profile of the eigenvector $A^i(\xi)$ differs strongly from a plane wave illumination, with most of the energy concentrated into the high order components. In Figure~\ref{fig:field_varyNA} the optimum solution is plotted for $\alpha=85^\circ$, $h=\lambda/5$ and $\text{NA}=0.001, 0.01, 0.1, 0.5$. In this case, changing the numerical aperture of the system does not influence much the profile of the solution.

\begin{figure}[htbp]
\centering
\fbox{\includegraphics[scale=0.35]{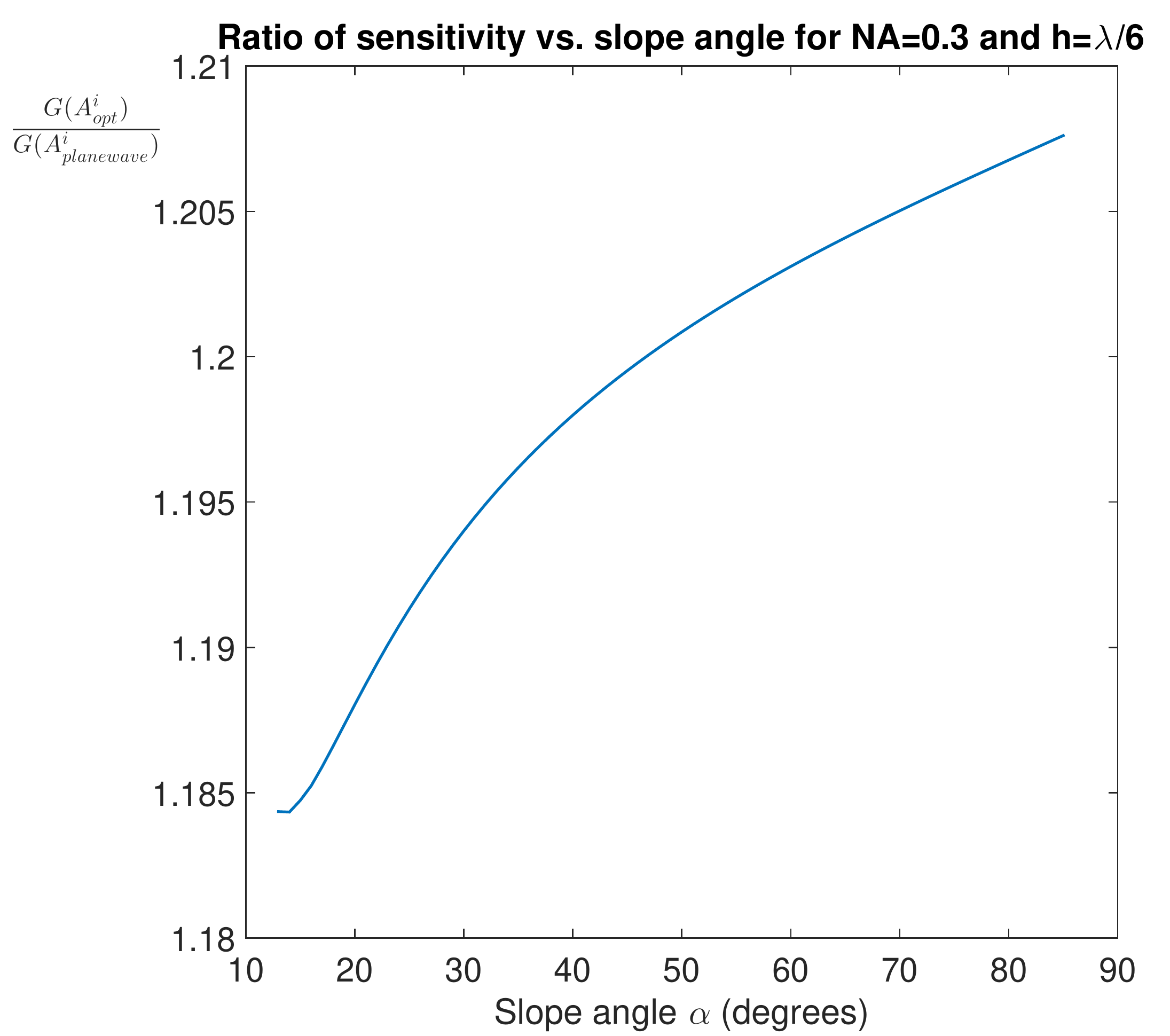}}
\caption{An optimized field (compare to a plane wave illumination) is much more sensitive to the side-wall angle of a cliff-like structure, given a fixed numerical aperture of the input pupil.}
\label{fig:sens_vs_angle}
\end{figure}

In Figure~\ref{fig:sens_vs_angle}, the sensitivity $G(A^i)$ for both the optimum field and the plane wave case is plotted against the variation of the slope angle $13^\circ\leq\alpha\leq 89^\circ$, for $\text{NA}=0.4$ and $h=\lambda/3$. Using the calculated solution gives a higher sensitivity over the whole range of investigated angles which increases for steep angles.

\begin{figure}[htbp]
\centering
\fbox{\includegraphics[scale=0.35]{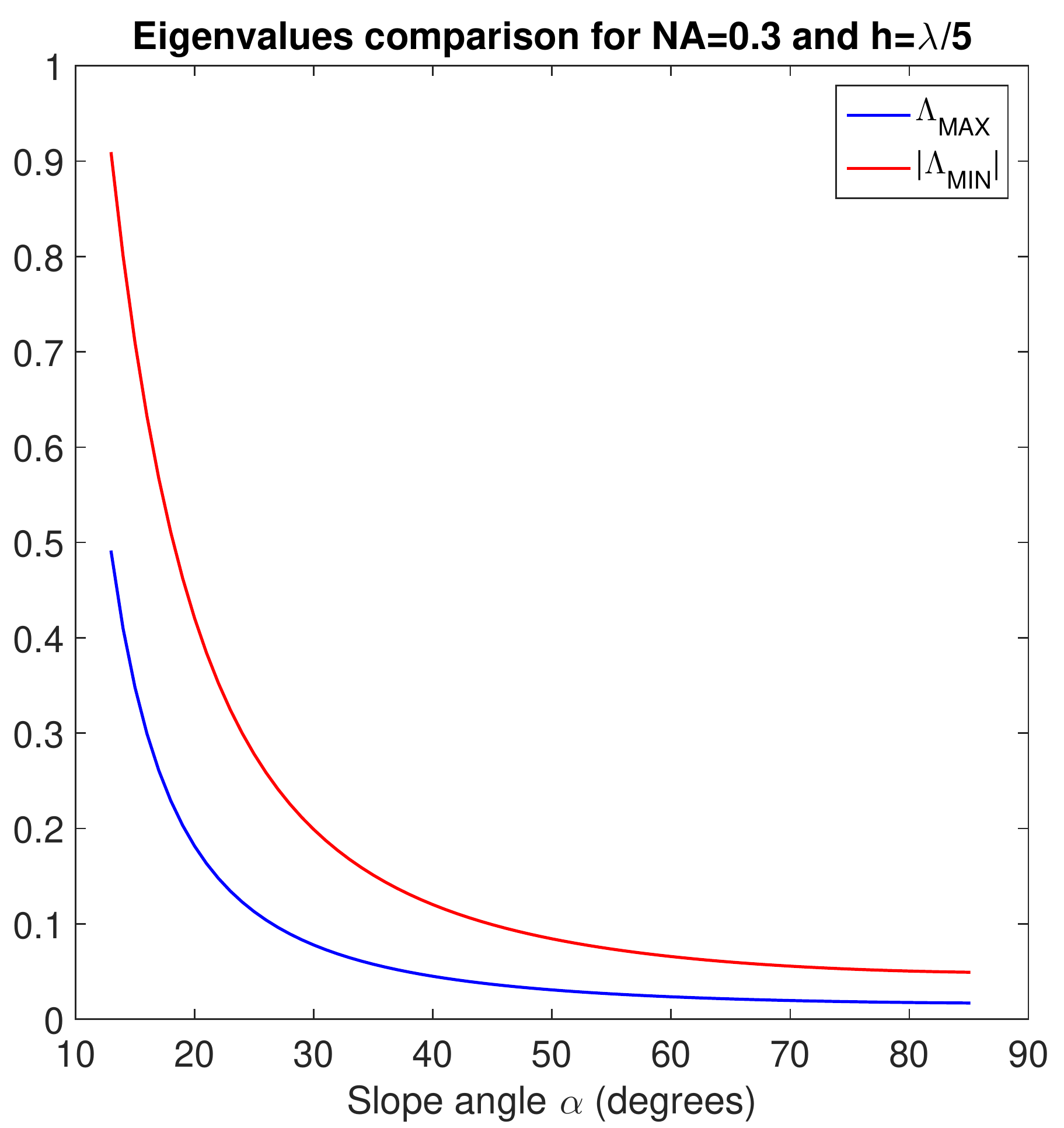}}
\caption{It is important to compute both the maximum eigenvalue and the minimum one, because the latter can also represent the best solution to the problem.}
\label{fig:compare_eigenvalues}
\end{figure}

It is important to point out that it is critical to find both the strongest positive and most negative eigenvalues of the system; in fact, we are interested in finding the input field which gives the steepest change in the reflected intensity. Hence its value could also diminish, thus the eigenvector associated to the strongest minimum could also be an acceptable solution of our system. This hypothesis is confirmed by Figure~\ref{fig:compare_eigenvalues}. Without any loss of generality, we chose to plot the values of the minimum and maximum eigenvalues for $\text{NA} = 0.3$ and $h=\lambda/5$, in this case we observe that the best solution is actually given by the strongest minimum eigenvalue. Therefore it is important to calculate and find both solutions.

\begin{figure}[htbp]
\centering
\fbox{\includegraphics[scale=0.35]{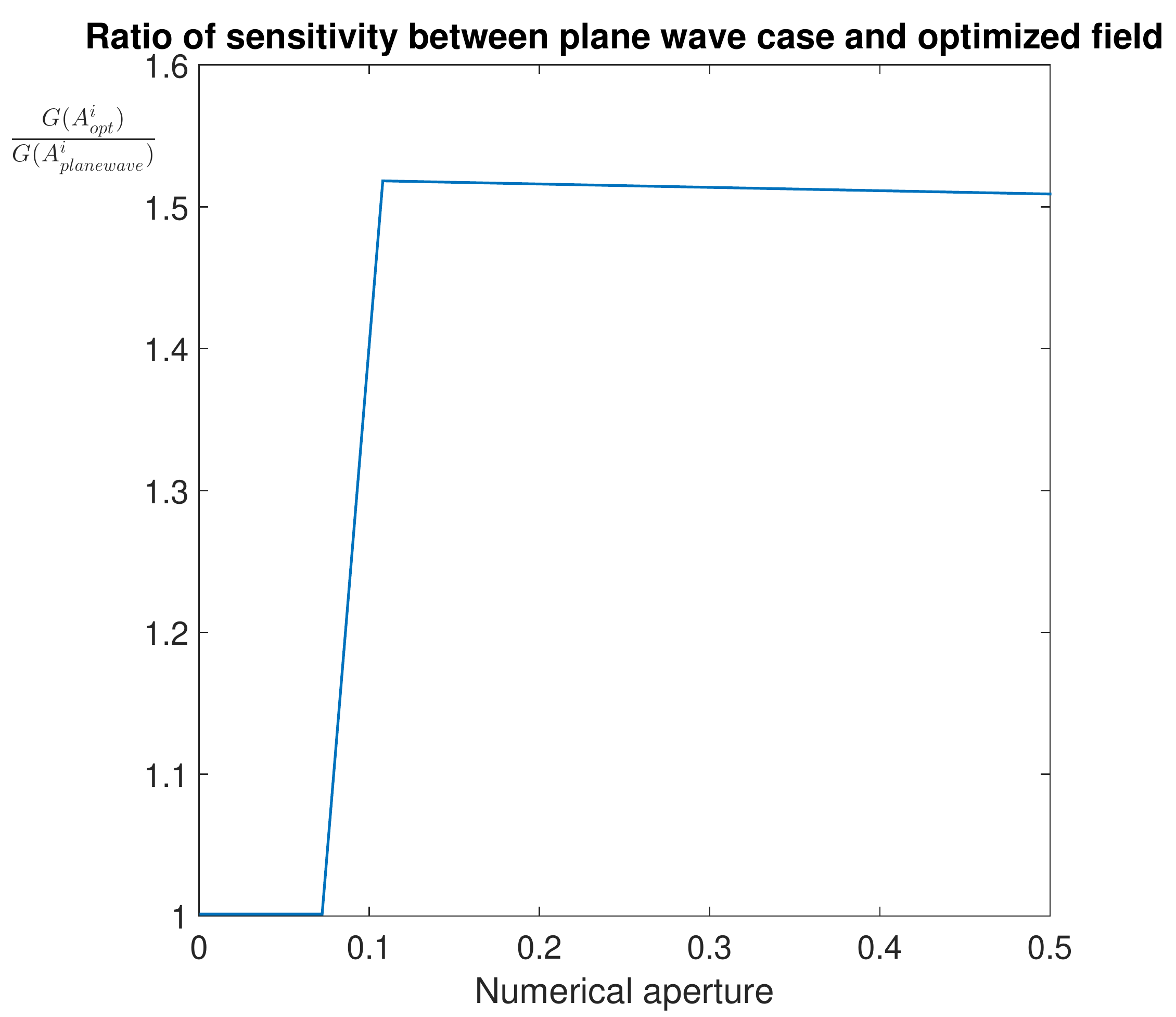}}
\caption{The ratio of the sensitivities approaches $1$ when $\text{NA}$ becomes small, as predicted from the theory.}
\label{fig:sens_vs_NA}
\end{figure}

Figure~\ref{fig:sens_vs_NA} shows the calculated value of the ratio of the functional $G(A^i)$ for the beam which provides the highest sensitivity and the value of $G$ for a plane wave illumination. For small values of the numerical aperture of the system, i.e. $\text{NA}\rightarrow 0$, we expect that the ratio converges to $1$. In this approximation, the eigenvalue of the optimum solution is given by Eq.~(\ref{eq:eigenvalue_limit_case}) in the Appendix, and the sensitivity for the plane wave case can be computed by taking the limit $\text{NA}\rightarrow 0$ in Eq.~(\ref{eq:sens_plane_int}). We have:
\begin{align}
G(A^i_{\text{PW}}) &= |A^i_0|^2\Re\int_{-\frac{\text{NA}}{\lambda}}^{+\frac{\text{NA}}{\lambda}}\int_{-\frac{\text{NA}}{\lambda}}^{+\frac{\text{NA}}{\lambda}} \mathcal{K}(\xi,\xi')\,d\xi\,d\xi'\notag\\
&=\frac{\lambda}{2\text{NA}}4\left(\frac{\text{NA}}{\lambda}\right)^2 \mathcal{K}(0,0)\notag\\
&=4\frac{\text{NA}}{\lambda}\frac{h k \cos(hk) - \sin(hk)}{k\sin^2\alpha}\,\cos(kh)
\end{align}
that is exactly the same expression reported in Eq.~(\ref{eq:max_sens}).

\section{Conclusion}\label{sec:conclusion}

In this paper we determined the illumination which is most sensitive to the change of the side-wall angle of a cliff-like structure. We expressed this problem mathematically by seeking the maximum (or minimum) of a given functional under the constraint in which the power carried by the beam is fixed. As a result, we have solved a constraint optimization problem that can be formulated in terms of a Lagrange multiplier rule. The optimum field was shown to be the eigenfunction associated with the strongest positive (or strongest negative) eigenvalue. The solution to this problem cannot be found analytically, therefore we used Gauss numerical integration to solve it. Nevertheless, for the limit case $\text{NA}\rightarrow 0$ is possible to find a closed formula for the optimum field, which can also be used to test our numerical routines. The improvement in the sensitivity detection is substantially higher compared to what we can achieve when we use plane wave illumination. These results emphasize how important it is to shape an optical beam according to the requirements imposed by experimental measurements or theoretical simulation; even in the presence of a rather simple model, where we are not dealing with periodic structures and we restrict ourselves to the scalar approximation, finding the optimized field $A^i(\xi)$ has proven to be a crucial element for the improvement in the detection of the side-wall angle.

\begin{figure}[htbp]
\centering
\fbox{\includegraphics[scale=0.35]{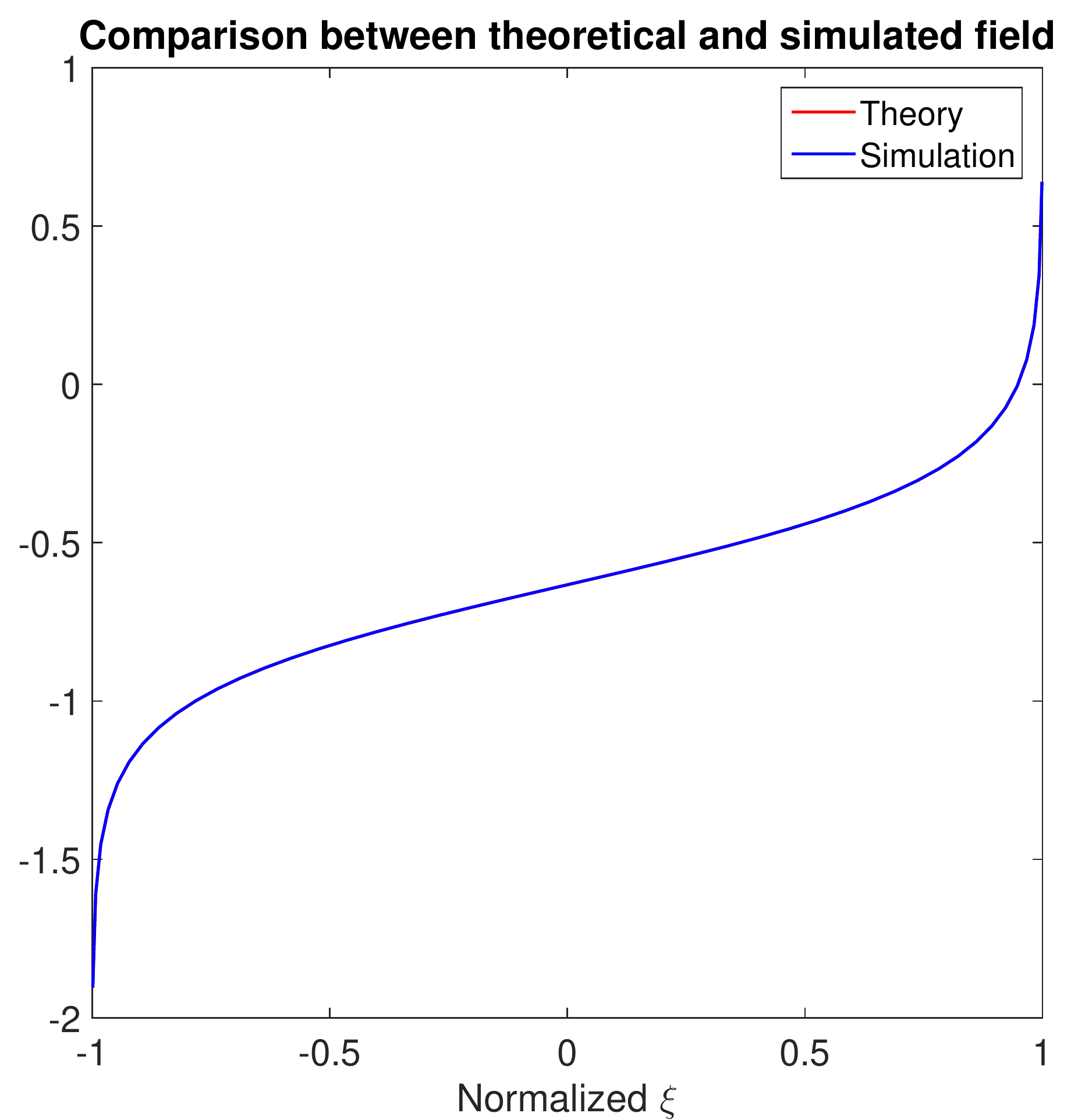}}
\caption{When we consider the case $\text{NA}\rightarrow 0$, there is a perfect agreement between simulation and theory.}
\label{fig:simul_plus_theory}
\end{figure}

\appendix

\section*{Appendix A: Solving the eigenvalue equation for the case \texorpdfstring{$NA\rightarrow 0$}{NA->0}}\label{app:A}
\setcounter{equation}{0}
\renewcommand{\theequation}{A{\arabic{equation}}}

We previously stated that Eq.~(\ref{eq:eigen_eq}) cannot be solved analytically and hence we used Gauss numerical integration to calculate the optimized pupil field $A^i(\xi)$. However, if we consider the limit $\text{NA}\rightarrow 0$, it is possible to reach a closed formula for the eigenvalue $\Lambda$ and its eigenfunction $A^i(\xi)$; in this appendix we will elucidate the mathematical derivation that leads to such a result. This result can also be used as a test of the numerical method for the general case.

First of all, the reader should notice that Eq~(\ref{eq:expression_H_limit}) is exact, meaning we did not calculated the results considering the limit $\text{NA}\rightarrow 0$, yet. If we do so, we see that the first integrand in the right hand-side of Eq.~(\ref{eq:expression_H_limit}) is $o(|\xi''-\xi'|)\leq \text{max}\,|\xi'|,|\xi''|$. Since $|\xi'|,|\xi''|<\text{NA}/\lambda$ the integrand is $o\left(\text{NA}/\lambda\right)$ and hence the integral is $o(\text{NA}^2/\lambda^2)$, while the last term is $o(\text{NA}/\lambda)$. Therefore the dominant terms of $\mathcal{H}(\xi,\xi')$ are:
\begin{align}
\mathcal{H}(\xi,\xi') \approx &\cos(kh) \frac{d\mathcal{F}(r_\alpha)}{d\alpha}(\xi'-\xi)\notag+\\
&- \frac{1}{\pi}\ln\left|\frac{\frac{\text{NA}}{\lambda}-\xi'}{\frac{\text{NA}}{\lambda}+\xi'}\right|\,\sin\left(k h \right)\,\frac{d\mathcal{F}(r_\alpha)}{d\alpha}(\xi'-\xi).\label{eq:approx_H}
\end{align}
Subsequently, we need to substitute Eq.~(\ref{eq:approx_H}) into Eq.~(\ref{eq:kernel_symmetric}) and solve Eq.~(\ref{eq:eigen_eq}) by considering, once again, the limit for $\text{NA}\rightarrow 0$; by retaining only the terms which substantially contribute to the final result, we are left with the following equation:
\begin{align}
2&\frac{\text{NA}}{\lambda}\biggl\{\cos(kh)\biggl[\frac{d\mathcal{F}(r_\alpha)}{d\alpha}(\xi) + \frac{d\mathcal{F}(r_\alpha)}{d\alpha}(-\xi)\biggr] + \notag\\
&-\frac{1}{\pi}\ln\left|\frac{\frac{\text{NA}}{\lambda}-\xi}{\frac{\text{NA}}{\lambda}+\xi}\right|\,\sin\left(k h \right)\,\frac{d\mathcal{F}(r_\alpha)}{d\alpha}(\xi)\biggr\}A^i(0) = \Lambda A^i(\xi)
\end{align}
We can find the eigenvalue $\Lambda$ by considering the case $\xi=0$:
\begin{align}
\Lambda &= 4\,\cos(kh)\frac{d\mathcal{F}(r_\alpha)}{d\alpha}(0)\,\frac{\text{NA}}{\lambda}\notag \\
&= 4\,\frac{\text{NA}}{\lambda}\frac{h k \cos^2(hk) - \cos(kh) \sin(hk)}{k\sin^2\alpha}\label{eq:eigenvalue_limit_case}
\end{align}
where we have substituted for:
\begin{equation}
\frac{d\mathcal{F}(r_\alpha)}{d\alpha}(0) = \frac{h k \cos(hk) - \sin(hk)}{k\sin^2\alpha}.
\end{equation}
Using the expression for the eigenvalue $\Lambda$, we can calculate the analytic expression for the eigenfunction $A^i(\xi)$:
\begin{align}
A^i(\xi) &= \frac{A^i(0)}{2}\biggl[\frac{d\mathcal{F}(r_\alpha)}{d\alpha}(\xi) + \frac{d\mathcal{F}(r_\alpha)}{d\alpha}(-\xi) + \notag\\
&-\frac{1}{\pi}\ln\left|\frac{\frac{\text{NA}}{\lambda}-\xi}{\frac{\text{NA}}{\lambda}+\xi}\right|\,\tan\left(k h \right)\,\frac{d\mathcal{F}(r_\alpha)}{d\alpha}(\xi)\biggr]\frac{1}{\frac{d\mathcal{F}(r_\alpha)}{d\alpha}(0)}\label{eq:final_Ai}
\end{align}
Finally, $A^i(0)$ follows from:
\begin{equation}
\int_{-\frac{\text{NA}}{\lambda}}^{+\frac{\text{NA}}{\lambda}} |A^i(\xi)|^2 \,d\xi = 1,
\end{equation}
The optimum sensitivity that can be achieved for a given $\alpha$ can be written as:
\begin{equation}
\Lambda = \frac{\delta G(A^i)(A^i)}{\delta P(A^i)(A^i)} = \frac{G(A^i)}{P(A^i)} = G(A^i) = \text{max}/\text{min}\,G(A^i)\label{eq:max_sens}
\end{equation} 
Hence:
\begin{equation}
\text{max}/\text{min}\,G(A^i) = \Lambda = 4\,\frac{\text{NA}}{\lambda}\frac{h k \cos^2(hk) - \cos(kh) \sin(hk)}{k\sin^2\alpha}
\end{equation}
we therefore notice that the sensitivity $G(A^i)$ decreases monotonically as the slope angle $\alpha$ increases but it increases with $\text{NA}$.

\section*{Appendix B: The Principal Value integral}\label{app:B}
\setcounter{equation}{0}
\renewcommand{\theequation}{B{\arabic{equation}}}

In this section we will perform the full demonstration regarding the integral containing the generalized function given by the Principal Value, used to be obtain an analytic expression of $\mathcal{H}(\xi,\xi')$ in Eq.~(\ref{eq:expression_H_limit}). Let $\phi$ be a small test function. Then:
\begin{align}
&\text{PV} \int_{-\frac{\text{NA}}{\lambda}}^{+\frac{\text{NA}}{\lambda}} \frac{\sin\left(k h + \pi h\frac{\xi}{\tan\alpha}\right)}{\pi(\xi-\xi')} \phi(\xi) d\xi \notag\\
&=\frac{1}{\pi} \lim_{\epsilon\rightarrow 0}\biggl[\int_{-\frac{\text{NA}}{\lambda}}^{\xi'-\epsilon} \frac{\sin\left(k h + \pi h\frac{\xi}{\tan\alpha}\right)}{\xi-\xi'}\,\phi(\xi)\;d\xi+\notag\\
&+\int_{\xi'+\epsilon}^{+\frac{\text{NA}}{\lambda}}\frac{\sin\left(k h + \pi h\frac{\xi}{\tan\alpha}\right)}{\xi-\xi'}\,\phi(\xi)\;d\xi\biggr]\label{eq:PV}
\end{align}
Since $\sin\left(k h + \pi h\xi/\tan\alpha\right)$ is a smooth function, $ \psi(\xi)=\sin\left(k h + \pi h\xi/\tan\alpha\right)\,\phi(\xi)$ is also smooth and hence we can write:
\begin{align}
&\text{PV} \int_{-\frac{\text{NA}}{\lambda}}^{+\frac{\text{NA}}{\lambda}} \frac{\sin\left(k h + \pi h\frac{\xi}{\tan\alpha}\right)}{\pi(\xi-\xi')} \phi(\xi) d\xi =\notag\\
&= \frac{1}{\pi} \lim_{\epsilon\rightarrow 0}\biggl[\int_{-\frac{\text{NA}}{\lambda}}^{\xi'-\epsilon} \frac{\psi(\xi)}{\xi-\xi'} d\xi + \int_{\xi'+\epsilon}^{+\frac{\text{NA}}{\lambda}}\frac{\psi(\xi)}{\xi-\xi'}\,d\xi\biggr]\notag\\
&= \frac{1}{\pi} \lim_{\epsilon\rightarrow 0}\biggl[\biggl(\int_{-\frac{\text{NA}}{\lambda}}^{\xi'-\epsilon}+\int_{\xi'+\epsilon}^{+\frac{\text{NA}}{\lambda}}\biggr) \frac{\psi(\xi)-\psi(\xi')}{\xi-\xi'} d\xi +\notag\\
&\;+\psi(\xi')\biggl(\int_{-\frac{\text{NA}}{\lambda}}^{\xi'-\epsilon}+\int_{\xi'+\epsilon}^{+\frac{\text{NA}}{\lambda}}\biggr)\frac{1}{\xi-\xi'}\,d\xi\biggr]\notag\\
&= \frac{1}{\pi} \biggl[\int_{-\frac{\text{NA}}{\lambda}}^{+\frac{\text{NA}}{\lambda}} \frac{\psi(\xi)-\psi(\xi')}{\xi-\xi'} d\xi +\notag\\
&\;+\psi(\xi')\lim_{\epsilon\rightarrow 0}\biggl(\int_{-\frac{\text{NA}}{\lambda}}^{\xi'-\epsilon}+\int_{\xi'+\epsilon}^{+\frac{\text{NA}}{\lambda}}\biggr)\frac{1}{\xi-\xi'}\,d\xi\biggr]\notag\\
&= \frac{1}{\pi} \int_{-\frac{\text{NA}}{\lambda}}^{+\frac{\text{NA}}{\lambda}} \frac{\psi(\xi)-\psi(\xi')}{\xi-\xi'} d\xi + \frac{\psi(\xi')}{\pi} \ln\left|\frac{\frac{\text{NA}}{\lambda}-\xi'}{\frac{\text{NA}}{\lambda}+\xi'}\right|.
\end{align}

\section*{Funding Information}

This work is supported by NanoNextNL, a micro and nanotechnology consortium of the Government of the Netherlands and 130 partners.

\bibliography{bibtex_scalar_optimization}

\end{document}